# Spin-Dependent Dokshitzer-Gribov-Lipatov- Altarelli-Parisi equations and t-evolution of structure functions in leading order at small-*x*

R. Rajkhowa

Physics Department, T.H.B. College, Jamugurihat, Sonitpur, Assam.

E-mail:rasna.rajkhowa@gmail.com

**Abstract**

In this paper the spin-dependent singlet and nonsinglet structure functions have been obtained by solving Dokshitzer-Gribov-Lipatov-Altarelli-Parisi evolution equations in leading order in the small $x$ limit. Here we have used Taylor series expansion and then the particular and unique solution to solve the evolution equations. We have also calculated $t$ evolutions of deuteron, proton and neutron structure functions and the results are compared with the SLAC E-143 Collaboration data.

## 1. Introduction

DIS of polarized electrons and muons off polarized targets has been used to study the internal spin structure of the nucleon. The most abundant and accurate experimental information we have so far comes from the so called longitudinal spin-dependent structure function $g_1$ which is obtained with longitudinally polarized leptons on longitudinally polarized protons, deuterons, and $^3He$ targets and it allows separate determination of spin-dependent deuteron, proton and neutron structure functions [1-8].

In the polarized deep-inelastic scattering (DIS), the spin structure of the nucleon has been studied by using polarized lepton beams scattered by polarized targets. These fixed-target experiments have been used to characterize the spin structure of the proton and neutron and to test additional fundamental QCD and quark-parton model (QPM) sum rules. The first experiments in polarized electron-polarized proton scattering, performed in the 1970s, helped establish the parton structure of the proton. In the late 1980s, a polarized muon-polarized proton experiment found that a QPM sum rule was violated, which seemed to indicate that the quarks do not account for the spin of the proton. This ''proton-spin crisis'' gave birth to a new generation of experiments at several high-energy physics laboratories around the world.

The new and extensive data sample collected from these fixed-target experiments has enabled a careful characterization of the spin dependent parton substructure of the nucleon. The results have been used to test QCD, to find an independent value for $\alpha_S(Q^2)$, and to probe with reasonable precision the polarized parton distributions. Recent interest in the spin structure of the proton, neutron, and deuteron and advances in experimental techniques have led to a number of experiments concerned with DIS of polarized leptons on various polarized targets. Among these are the E143 experiments at SLAC [9] and those of the SMC Collaboration at CERN [10], which used polarized hydrogen and deuterium; the E154 experiment at SLAC [11] and the HERMES Collaboration experiments at DESY [12], which used polarized $^3$He; and the HERMES experiment [13], which used polarized hydrogen [14]. A new material, deuterized lithium $^6LiD$, has recently emerged as a source of polarized deuterium in the E155/E155x experiments at SLAC [15]. The spin-dependent structure function $g_1(x, Q^2)$ for deep-inelastic leptonnucleon scattering is of fundamental importance in understanding the quark and gluon spin structure of the proton and neutron. According to the DGLAP equations [16], $g_1(x, Q^2)$ is expected to evolve logarithmically with $Q^2$, where $g_1$ depends both on $x$, the fractional momentum carried by the struck parton, and on $Q^2$, the squared fourmomentum of the exchanged virtual photon. There have been a number of theoretical approaches [17, 18] to calculate $g_1(x, Q^2)$ using phenomenological models of nucleon structure.

The present paper reports particular and unique solutions of polarized DGLAP evolution equations computed from complete solutions in leading order at low-$x$ and calculation of $t$ evolutions for singlet, non-singlet, structure functions and hence $t$-evolutions of deuteron, proton, neutron structure functions. Here, the integro-differential polarized DGLAP evolution equations have been converted into first order partial differential equations by applying Taylor expansion in the small-$x$ limit. Then they have been solved by standard analytical methods. The results of $t$-evolutions are compared with the SLAC E-143 Collaboration data.

In the present paper, section 1 is the introduction. In section 2 necessary theory has been discussed. Section 3 gives results and discussion, and section 4 is conclusion.

## 2.    Theory

When both the beam and the target are longitudinally polarized in DIS, the asymmetry is defined as

$$A_\| = \frac{\sigma^{\uparrow\downarrow} - \sigma^{\uparrow\uparrow}}{\sigma^{\uparrow\downarrow} + \sigma^{\uparrow\uparrow}}$$

where $\sigma^{\uparrow\downarrow}$ and $\sigma^{\uparrow\uparrow}$ are the cross sections for the opposite and same spin directions, respectively. Similarly, the transverse asymmetry, determined from scattering of a longitudinally polarized beam on a transversely polarized target, is defined as

$$A_\perp = \frac{\sigma^{\Downarrow\rightarrow} - \sigma^{\Uparrow\rightarrow}}{\sigma^{\Downarrow\rightarrow} + \sigma^{\Uparrow\rightarrow}}$$

These asymmetries can be express in terms of longitudinal ($A_1$) and transverse ($A_2$) virtual photon-nucleon asymmetries as

$$A_\| = D[A_1 + \eta A_2] \quad \text{and} \quad A_\perp = d[A_2 - \eta\gamma\left(1 - \frac{y}{2}\right)A_1]$$

where

$$D = \frac{2y - y^2}{2(1-y)(1+R) + y^2}, \quad \eta = \left(\frac{Q}{E}\right)\frac{2(1-y)}{y(2-y)}, \quad d = \frac{\sqrt{1-y}}{1-y/2}D, \quad \gamma^2 = 4y^2 x^2 \quad \text{and} \quad y = \frac{M}{Q}$$

The virtual photon-nucleon asymmetries for the proton, neutron, and deuteron are defined as

$$A_1^{p,n} = \frac{\sigma_{1/2} - \sigma_{3/2}}{\sigma_{1/2} + \sigma_{3/2}}, \quad A_2^{p,n} = \frac{2\sigma^{TL}}{\sigma_{1/2} + \sigma_{3/2}}, \quad A_1^d = \frac{\sigma_0 - \sigma_2}{\sigma_0 + \sigma_2} \quad \text{and} \quad A_2^d = \frac{\sigma_0^{TL} - \sigma_1^{TL}}{\sigma_{1/2} + \sigma_{3/2}}$$

The longitudinal spin-dependent structure function $g_1(x)$ is defined as

$$g_1(x) = \frac{1}{2}\sum e_i^2 \Delta q_i(x)$$

where

$$q_i(x) = q_i^+ + q_i^{-+}(x) - q_i^-(x) + q_i^{--}(x)$$

Here $q_i^+$ and $q_i^-(x)$ are the densities of quarks of flavor "$i$" with helicity parallel and antiparallel to the nucleon spin. The spin-dependent structure functions $g_1(x, Q^2)$ and $g_2(x, Q^2)$ are related to the spin-independent structure function $F_2(x, Q^2)$ as

$$g_1 = \frac{F_2(x,Q^2)[A_1(x,Q^2) + \gamma A_2(x,Q^2)]}{2x[1 + R_\sigma(x,Q^2)]} \quad \text{and} \quad g_2 = \frac{F_2(x,Q^2)[-A_1(x,Q^2) + \frac{A_2(x,Q^2)}{\gamma}]}{2x[1 + R_\sigma(x,Q^2)]}$$

where $R_\sigma = \frac{\sigma_L(x,Q^2)}{\sigma_T(x,Q^2)}$ is the ratio of the longitudinal and transverse virtual photon cross sections.

The polarized DGLAP evolution equation [16] in the standard form is given by

$$\frac{\partial g_1(x,Q^2)}{\partial \ln Q^2} = P(x,Q^2) \otimes g_1(x,Q^2)$$

where $g_1(x, Q^2)$ is the spin-dependent structure function as a function of $x$ and $Q^2$, where $x$ is the Bjorken variable and $Q^2$ is the four-momentum transfer in a DIS process. Here $P(x, Q^2)$ is the spin-dependent kernel known perturbatively up to the first few orders in $\alpha_s(Q^2)$, the strong coupling constant. Here $\otimes$ represents the standard Mellin convolution, and the notation is given by

$$a(x) \otimes b(x) = \int_0^1 \frac{dy}{y} a(y) b\left(\frac{x}{y}\right)$$

One can write

$$P(x, Q^2) = \frac{\alpha_s(Q^2)}{2\pi} P^{(0)}(x) + \left(\frac{\alpha_s(Q^2)}{2\pi}\right)^2 P^{(1)}(x)$$

where $P^{(0)}(x)$ and $P^{(1)}(x)$ are spin-dependent splitting functions in LO and NLO.

The singlet and non-singlet structure functions [8, 19, 20] are obtained from the polarized DGLAP evolution equations as

$$* \frac{\partial g_1^S(x,t)}{\partial t} - \frac{\alpha_s'(t)}{2\pi} [\frac{2}{3}\{3 + 4\ln(1-x)\} g_1^S(x,t) + \frac{4}{3}\int_x^1 \frac{dw}{1-w}\{(1+w^2) g_1^S\left(\frac{x}{w}, t\right) - 2g_1^S(x,t)\}]$$

$$+ n_f \int_x^1 (2w-1)\Delta G(\frac{x}{w}, t) dw\}] = 0, \qquad (1)$$

$$* \frac{\partial g_1^{NS}(x,t)}{\partial t} - \frac{\alpha_s'(t)}{2\pi} [\frac{2}{3}\{3 + 4\ln(1-x)\} g_1^{NS}(x,t) + \frac{4}{3}\int_x^1 \frac{dw}{1-w}\{(1+w^2) g_1^{NS}\left(\frac{x}{w}, t\right) - 2g_1^{NS}(x,t)\}] = 0, \qquad (2)$$

in LO.

Let us introduce the variable $u = 1-w$ and note that [21]

$$\frac{x}{w} = \frac{x}{1-u} = x + \frac{xu}{1-u}. \qquad (3)$$

The series (3) is convergent for $|u| < 1$. Since $x < w < 1$, so $0 < u < 1-x$ and hence the convergence criterion is satisfied. Now, using Taylor expansion method [22, 23] we can rewrite $g_1^S(x/w, t)$ as

$$g_1^S(x/w, t) = g_1^S\left((x + \frac{xu}{1-u}), t\right)$$

$$= g_1^S(x,t) + x \cdot \frac{u}{1-u} \frac{\partial g_1^S(x,t)}{\partial x} + \frac{1}{2} x^2 \left(\frac{u}{1-u}\right)^2 \frac{\partial^2 g_1^S(x,t)}{\partial x^2} + \ldots$$

which covers the whole range of $u$, $0 < u < 1-x$. Since $x$ is small in our region of discussion, the terms containing $x^2$ and higher powers of $x$ can be neglected as our first approximation as discussed in our earlier works [24-26]. $g_1^S(x/w,t)$ can then be approximated for small-$x$ as

$$g_1^S(x/w,t) \cong g_1^S(x,t) + x \cdot \frac{u}{1-u} \frac{\partial g_1^S(x,t)}{\partial x}. \tag{4}$$

Similarly, $G(x/w, t)$ can be approximated for small-$x$ as

$$\Delta G(x/w,t) \cong \Delta G(x,t) + x \cdot \frac{u}{1-u} \frac{\partial \Delta G(x,t)}{\partial x}. \tag{5}$$

Using equations (4) and (5) in equation (1) and performing $u$-integrations we get

$$\frac{\partial g_1^S(x,t)}{\partial t} - \frac{\alpha_s'(t)}{2\pi}\left[A_1(x)g_1^S(x,t) + A_2(x)\Delta G(x,t) + A_3(x)\frac{\partial g_1^S(x,t)}{\partial x} + A_4(x)\frac{\partial \Delta G(x,t)}{\partial x}\right] = 0. \tag{6}$$

Here

$$A_1(x) = \frac{2}{3}\{3 + 4\ln(1-x) + (x-1)(x+3)\}, \quad A_2(x) = N_f x(1-x),$$

$$A_3(x) = \frac{2}{3}\{x(1-x^2) + 2x\ln(\frac{1}{x})\}, \quad A_4(x) = N_f x\{(1-x)(2-x) + \ln x\}.$$

We assume [24-26, 28]

$$\Delta G(x,t) = K(x) g_1^S(x,t). \tag{7}$$

Here, $K$ is a function of $x$. It is to be noted that if we consider Regge behaviour of singlet and gluon structure function, it is possible to solve coupled evolution equations for singlet and gluon structure functions and evaluate $K(x)$ in LO and NLO. Otherwise this is a parameter to be estimated from experimental data. We take $K(x) = k$, $ax^b$, $ce^{-dx}$, where $k, a, b, c, d$ are constants. Therefore equations (6) becomes

$$\frac{\partial g_1^S(x,t)}{\partial t} - \frac{A_f}{t}\left[L_1(x)g_1^S(x,t) + L_2(x)\frac{\partial g_1^S(x,t)}{\partial x}\right] = 0. \tag{8}$$

Here,

$$L_1(x) = A_1(x) + K(x)A_2(x) + A_4(x)\frac{\partial K(x)}{\partial x}, \quad L_2(x) = A_3(x) + K(x)A_4(x) \text{ and } A_f = 4/(33-2n_f).$$

The general solution [23, 27] of equation (8) is $g(U, V) = 0$, where $g$ is an arbitrary function and $U(x, t, g_1^S) = C_1$ and $V(x, t, g_1^S) = C_2$, where $C_1$ and $C_2$ are constants and they form a solutions of equations

$$\frac{dx}{A_f L_2(x)} = \frac{dt}{-t} = \frac{dg_1^S(x,t)}{-A_f L_1(x)g_1^S(x,t)}. \tag{9}$$

Solving equation (9) we obtain

$$U\left(x,t,g_1^S\right) = t\exp\left[\frac{1}{A_f}\int\frac{1}{L_2(x)}dx\right] \text{ and } V\left(x,t,g_1^S\right) = g_1^S(x,t)\exp\left[\int\frac{L_1(x)}{L_2(x)}dx\right].$$

## 2. (a) Complete and Particular Solutions

Since $U$ and $V$ are two independent solutions of equation (9) and if $\alpha$ and $\beta$ are arbitrary constants, then $V = \alpha U + \beta$ may be taken as a complete solution [23, 27] of equation (8). We take this form as this is the simplest form of a complete solution which contains both the arbitrary constants $\alpha$ and $\beta$. Earlier [28] we considered a solution $AU + BV = 0$, where $A$ and $B$ are arbitrary constants. But that is not a complete solution having both the arbitrary constants as this equation can be transformed to the form $V = CU$, where $C = -A/B$, i. e, the equation contains only one arbitrary constant. So, the complete solution

$$g_1^S(x,t)\exp\left[\int\frac{L_1(x)}{L_2(x)}dx\right] = \alpha t\exp\left[\frac{1}{A_f}\int\frac{L_1(x)}{L_2(x)}dx\right] + \beta \tag{10}$$

is a two-parameter family of surfaces, which does not have an envelope, since the arbitrary constants enter linearly [23,27]. Differentiating equation (10) with respect to $\beta$ we get $0 = 1$, which is absurd. Hence there is no singular solution. The one parameter family determined by taking $\beta = \alpha^2$ has equation

$$g_1^S(x,t)\exp\left[\int\frac{L_1(x)}{l_2(x)}dx\right] = \alpha t\exp\left[\frac{1}{A_f}\int\frac{1}{L_2(x)}dx\right] + \alpha^2. \tag{11}$$

Differentiating equation (11) with respect to $\alpha$, we get $\alpha = -\frac{1}{2}t\exp\left[\frac{1}{A_f}\int\frac{1}{L_2(x)}dx\right]$. Putting the value of $\alpha$ in equation (11), we get

$$g_1^S(x,t) = -\frac{1}{4}t^2\exp\left[\int\left(\frac{2}{A_f L_2(x)} - \frac{L_1(x)}{L_2(x)}\right)dx\right], \tag{12}$$

which is merely a particular solution of the general solution. Now, defining

$$g_1^S(x,t_0) = -\frac{1}{4}t_0^2\exp\left[\int\left(\frac{2}{A_f L_2(x)} - \frac{L_1(x)}{L_2(x)}\right)dx\right], \text{ at } t = t_0, \text{ where, } t_0 = \ln(Q_0^2/\Lambda^2) \text{ at any lower value}$$

$Q = Q_0$, we get from equation (12)

$$g_1^S(x,t) = g_1^S(x,t_0)\left(\frac{t}{t_0}\right)^2, \tag{13}$$

which gives the $t$-evolution of singlet structure function $g_1^S(x,t)$. Again defining,

$$g_1^S(x_0,t) = -\frac{1}{4}t^2 \exp\left[\int\left(\frac{2}{A_f L_2(x)} - \frac{L_1(x)}{L_2(x)}\right)dx\right]_{x=x_0},$$

we obtain from equation (12)

$$g_1^S(x,t) = g_1^S(x_0,t)\exp\left[\int_{x_0}^{x}\left(\frac{2}{A_f L_2(x)} - \frac{L_1(x)}{L_2(x)}\right)dx\right] \quad (14)$$

which gives the *x*-evolution of singlet structure function $g_1^S(x,t)$. Proceeding in the same way, we get

$$g_1^{NS}(x,t) = g_1^{NS}(x_0,t)\left(\frac{t}{t_0}\right)^2. \quad (15)$$

which give the *t*-evolutions of non-singlet structure functions in LO.

$$g_1^{NS}(x,t) = g_1^{NS}(x_0,t)\exp\left[\int_{x_0}^{x}\left(\frac{2}{A_f Q(x)} - \frac{P(x)}{Q(x)}\right)dx\right], \quad (16)$$

which give the *x*-evolutions of non-singlet structure functions in LO.

For all these particular solutions, we take $\beta = \alpha^2$. But if we take $\beta = \alpha$ and differentiate with respect to $\alpha$ as before, we can not determine the value of $\alpha$. In general, if we take $\beta = \alpha^y$, we get in the solutions, the powers of ($t/t_0$) and the numerators of the first term inside the integral sign be $y/(y-1)$ for *t* and *x*-evolutions respectively in LO.

For phenomenological analysis, we compare our results with various experimental structure functions. Deuteron, proton and neutron structure functions can be written as

$$g_1^d(x,t) = \frac{5}{9}g_1^S(x,t), \quad (17)$$

$$g_1^p(x,t) = \left[\frac{5}{18}g_1^S(x,t) + \frac{3}{18}g_1^{NS}(x,t)\right], \quad (18)$$

$$g_1^n(x,t) = \left[\frac{5}{18}g_1^S(x,t) - \frac{3}{18}g_1^{NS}(x,t)\right]. \quad (19)$$

Now using equations (13) (14), (15) in equations (17), (18) and (19) we will get *t*-evolutions of deuteron, proton, neutron and *x*-evolution of deuteron structure functions at low-*x* as

$$g_1^{d,p,n}(x,t) = g_1^{d,p,n}(x,t_0)\left(\frac{t}{t_0}\right)^2, \quad (20)$$

$$g_1^d(x,t) = g_1^d(x_0,t)\exp\left[\int_{x_0}^{x}\left(\frac{2}{A_f L_2(x)} - \frac{L_1(x)}{L_2(x)}\right)dx\right] \quad (21)$$

in LO for $\beta = \alpha^2$.

The determination of *x*-evolutions of proton and neutron structure functions like that of deuteron structure function is not suitable by this methodology; because to extract the *x*-evolution of proton and neutron structure functions, we are to put equations (14) and (16) in equations (18) and (19). But as the functions inside the integral sign of equations (14) and (16) are different, we need to separate the input functions $g_1^S(x_0,t)$ and $g_1^{NS}(x_0,t)$ from the data points to extract the *x*-evolutions of the proton and neutron structure functions, which may contain large errors.

## 2. (b) Unique Solutions

Due to conservation of the electromagnetic current, $g_1$ must vanish as $Q^2$ goes to zero [29, 30]. Also $R \to 0$ in this limit. Here *R* indicates ratio of longitudinal and transverse cross-sections of virtual photon in DIS process. This implies that scaling should not be a valid concept in the region of very low-$Q^2$. The exchanged photon is then almost real and the close similarity of real photonic and hadronic interactions justifies the use of the Vector Meson Dominance (VMD) concept [31-32] for the description of $F_2$. In the language of perturbation theory, this concept is equivalent to a statement that a physical photon spends part of its time as a 'bare', point-like photon and part as a virtual hadron [30]. The power and beauty of explaining scaling violations with field theoretic methods (i.e., radiative corrections in QCD) remains, however, unchallenged in as much as they provide us with a framework for the whole *x*-region with essentially only one free parameter $\Lambda$ [33]. For $Q^2$ values much larger than $\Lambda^2$, the effective coupling is small and a perturbative description in terms of quarks and gluons interacting weakly makes sense. For $Q^2$ of order $\Lambda^2$, the effective coupling is infinite and we cannot make such a picture, since quarks and gluons will arrange themselves into strongly bound clusters, namely, hadrons [29] and so the perturbation series breaks down at small-$Q^2$ [29]. Thus, it can be thought of $\Lambda$ as marking the boundary between a world of quasi-free quarks and gluons, and the world of pions, protons, and so on. The value of $\Lambda$ is not predicted by the theory; it is a free parameter to be determined from experiment. It should expect that it is of the order of a typical hadronic mass [29]. Since the value of $\Lambda$ is so small we can take at $Q = \Lambda$, $g_1^S(x,t) = 0$ due to conservation of the electromagnetic current [30]. This dynamical prediction agrees with most adhoc parameterizations and with the data [33]. Using this boundary condition in equation (10) we get $\beta = 0$ and

$$g_1^S(x,t) = \alpha t \exp\left[\int\left(\frac{1}{A_f L_2(x)} - \frac{L_1(x)}{L_2(x)}\right)dx\right]. \tag{22}$$

Now, defining $g_1^S(x,t_0) = \alpha t_0 \exp\left[\int\left(\frac{1}{A_f L_2(x)} - \frac{L_1(x)}{L_2(x)}\right)dx\right]$, at $t = t_0$, where $t_0 = \ln(Q_0^2/\Lambda^2)$ at any lower value $Q = Q_0$, we get from equations (22)

$$g_1^S(x,t) = g_1^S(x,t_0)\left(\frac{t}{t_0}\right), \tag{23}$$

which gives the *t*-evolutions of singlet structure function $g_1^S(x,t)$ in LO. Proceeding in the same way we get

$$g_1^S(x,t) = g_1^S(x_0,t)\exp\left[\int_{x_0}^{x}\left(\frac{1}{A_f L_2(x)} - \frac{L_1(x)}{L_2(x)}\right)dx\right] \tag{24}$$

which gives the *x*-evolutions of singlet structure function $F_2^S(x,t)$ in LO. Similarly, we get for non-singlet structure functions

$$g_1^{NS}(x,t) = g_1^{NS}(x,t_0)\left(\frac{t}{t_0}\right). \tag{25}$$

$$g_1^{NS}(x,t) = g_1^{NS}(x_0,t)\exp\left[\int_{x_0}^{x}\left(\frac{1}{A_f Q(x)} - \frac{P(x)}{Q(x)}\right)dx\right], \tag{26}$$

which give the *t* and *x*-evolutions of non-singlet structure functions in LO.

Therefore corresponding results for *t*-evolution of deuteron, proton, neutron structure functions and *x*-evolution of deuteron structure function are

$$g_1^{d,p,n}(x,t) = g_1^{d,p,n}(x,t_0)\left(\frac{t}{t_0}\right), \tag{27}$$

$$g_1^d(x,t) = g_1^d(x_0,t)\exp\left[\int_{x_0}^{x}\left(\frac{1}{A_f L_2(x)} - \frac{L_1(x)}{L_2(x)}\right)dx\right] \tag{28}$$

in LO.

Already we have mentioned that the determination of *x*-evolutions of proton and neutron structure functions like that of deuteron structure function is not suitable by this methodology. It is to be noted that unique solutions of evolution equations of different structure functions are same with particular solutions for *y* maximum ($y = \infty$) in $\beta = \alpha^y$ relation. The procedure we follow is to begin with input distributions inferred from experiment and to integrate the evolution equations (24) and (26) numerically.

## 3. Results and Discussion

In the present paper, we have compared the results of $t$-evolutions of spin-dependent deuteron, proton and neutron structure functions in LO with different experimental data sets measured by the SLAC-E-143 [34] collaboration. The SLAC-E-143 collaborations data sets give the measurement of the spin-dependent structure function of deuteron, proton and neutron in deep inelastic scattering of spin-dependent electrons at incident energies of 9.7, 16.2 and 29.1 GeV on a spin-dependent Ammonia target. Data cover the kinematical x range 0.024 to 0.75 and $Q^2$-range from 0.5 to 10 GeV$^2$.

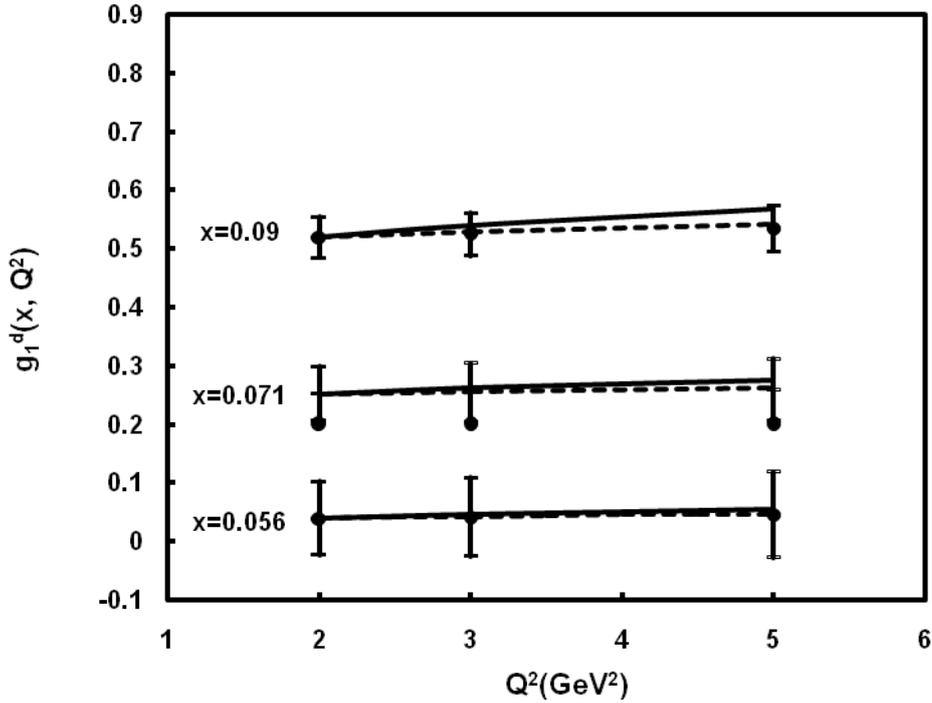

Fig.1

In fig.1, we present our results of $t$-evolutions of spin-dependent deuteron structure function for the representative values of $x$ given in the figures for $y = 2$ (solid lines) and $y$ maximum (dashed lines) in $\beta = \alpha^y$ relation in LO. Data points at lowest-$Q^2$ values in the figures are taken as input to test the evolution equation. Agreement with the data [34] is good.

In fig.2, we present our results of $t$-evolutions of spin-dependent proton structure function for the representative values of $x$ given in the figures for $y = 2$ (solid lines) and $y$ maximum (dashed lines) in $\beta = \alpha^y$ relation in LO. Data points at lowest-$Q^2$ values in the figures are taken as input to test the evolution equation. Agreement is found to be excellent.

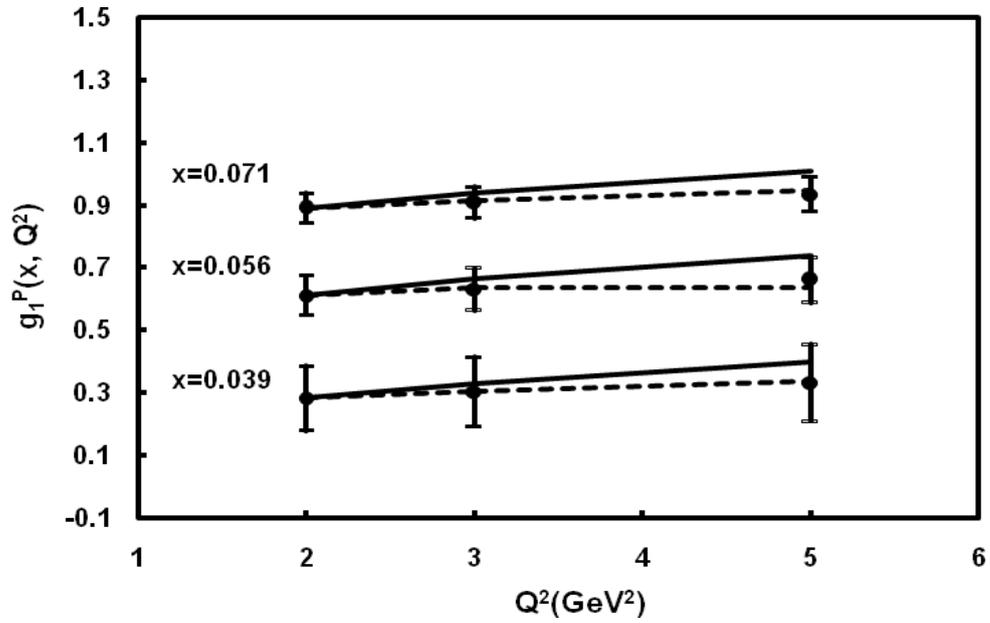

Fig.2

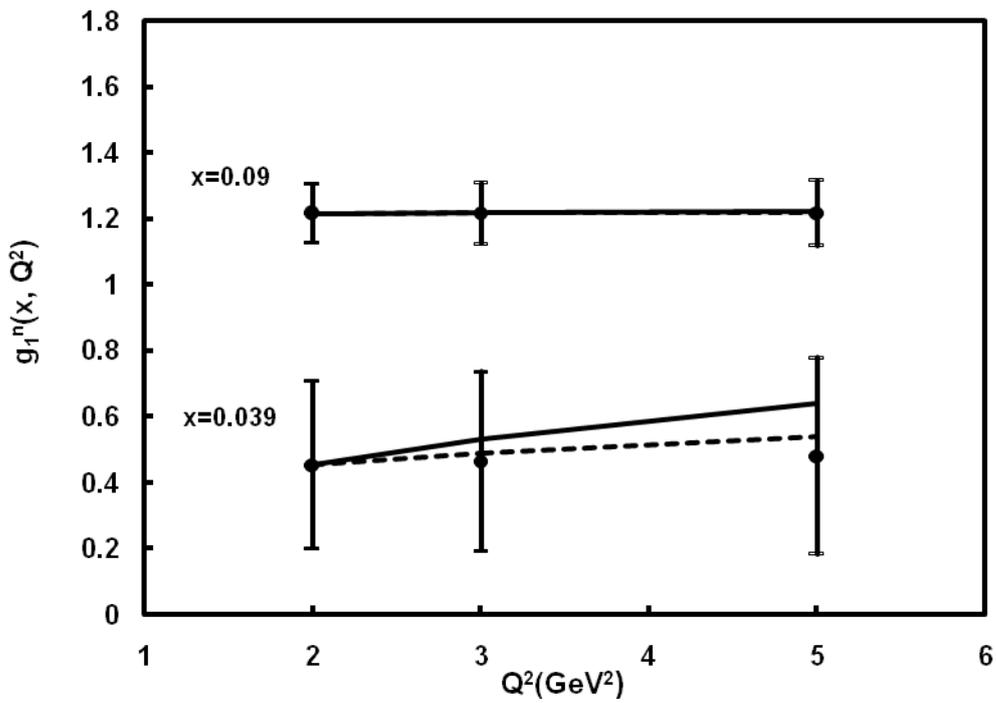

Fig.3

In fig.3, we present our results of *t*-evolutions of spin-dependent neutron structure function deuteron, proton for the representative values of *x* given in the figures for $y = 2$ (solid lines) and *y* maximum (dashed) in $\beta = \alpha^y$ relation in LO. Data points at lowest-$Q^2$ values in the figures are taken as input to test the evolution equation. Agreement with the data [34] is good.

Unique solutions of *t*-evolution for structure functions are same with particular solutions for *y* maximum ($y = \infty$) in $\beta = \alpha^y$ relation in LO.

## 4. Conclusion

We solve spin dependent DGLAP evolution equation in LO using Taylor expansion method and derive *t*-evolutions of various spin dependent structure functions and compare them with global data with satisfactory phenomenological success. It has been observed that though we have derived a unique *t*-evolution for deuteron, proton, neutron structure functions in LO, yet we can not establish a completely unique *x*-evolution for deuteron structure function in LO due to the relation $K(x)$ between singlet and gluon structure functions. $K(x)$ may be in the forms of a constant, an exponential function or a power function and they can equally produce required *x*-distribution of deuteron structure functions. But unlike many parameter arbitrary input *x*-distribution functions generally used in the literature, our method requires only one or two such parameters. On the other hand, we observed that the Taylor expansion method is mathematically simpler in comparison with other methods available in the literature. Explicit form of $K(x)$ can actually be obtained only by solving coupled DGLAP evolution equations for singlet and gluon structure functions. Though we study LO evolution equation for spin structure function, we hope that it can be extendable to NLO also. So we see that this simple method may have a wide application in solving DGLAP evolution equations.

**Figure Captions**

**Fig.1:** Results of *t*-evolutions of deuteron structure functions (solid lines for $y = 2$ and dashed lines for y maximum in $\beta = \alpha^y$ relation) for the representative values of *x* in LO for SLAC-E-143 data. For convenience, value of each data point is increased by adding 0.2*i*, where $i = 0, 1, 2$ are the numberings of curves counting from the bottom of the lowermost curve as the 0-th order. Data points at lowest-$Q^2$ values in the figures are taken as input.

**Fig.2:** Results of *t*-evolutions of poton structure functions (solid lines for $y = 2$ and dashed lines for y maximum in $\beta = \alpha^y$ relation) for the representative values of *x* in LO for SLAC-E-

143 data. For convenience, value of each data point is increased by adding $0.3i$, where $i = 0$, 1, 2 are the numberings of curves counting from the bottom of the lowermost curve as the 0-th order. Data points at lowest-$Q^2$ values in the figures are taken as input.

**Fig.3:** Results of *t*-evolutions of neutron structure functions (solid lines for $y = 2$ and dashed lines for y maximum in $\beta = \alpha^y$ relation) for the representative values of *x* in LO for SLAC-E-143 data. For convenience, value of each data point is increased by adding $0.3i$, where $i = 1, 4$ are the numberings of curves counting from the bottom of the lowermost curve as the 1st order. Data points at lowest-$Q^2$ values in the figures are taken as input.